\documentclass[conference]{llncs}
\usepackage{cite}
\usepackage{proof-at-the-end}
 
\usepackage{amsmath,amssymb,amsfonts,amsthm}
\usepackage{textcomp}
\usepackage{xcolor}

\def\BibTeX{{\rm B\kern-.05em{\sc i\kern-.025em b}\kern-.08em
    T\kern-.1667em\lower.7ex\hbox{E}\kern-.125emX}}
\usepackage{nicefrac}
\usepackage{siunitx}
\usepackage{array,framed}
\usepackage{amsthm}
\usepackage{hyperref}
\usepackage{booktabs}
\usepackage{appendix}

\newtheorem{manualtheoreminner}{Theorem}
\newenvironment{manualtheorem}[1]{%
  \renewcommand\themanualtheoreminner{#1}%
  \manualtheoreminner
}{\endmanualtheoreminner}

\usepackage{
  color,
  float,
  epsfig,
  wrapfig,
  graphics,
  graphicx,
  subcaption,
  verbatim
}
\usepackage{setspace}
\usepackage{latexsym,fancyhdr,url}
\usepackage{enumerate}
\usepackage{algorithm2e}
\usepackage{algpseudocode}
\usepackage{graphics}
\usepackage{xparse} 
\usepackage{xspace}
\usepackage{multirow}
\usepackage{csvsimple}
\usepackage{balance}

\usepackage{
  tikz,
  pgfplots,
  pgfplotstable
}
\usepackage{hyperref}

\usetikzlibrary{
  shapes.geometric,
  arrows,
  external,
  pgfplots.groupplots,
  matrix
}

\usepackage{fancyhdr}
\pgfplotsset{compat=1.9}

\usepackage{authblk}

\usepackage{mathtools}

\DeclareMathAlphabet{\mathcal}{OMS}{cmsy}{m}{n}

\DeclareGraphicsExtensions{%
    .png,.PNG,%
    .pdf,.PDF,%
    .jpg,.mps,.jpeg,.jbig2,.jb2,.JPG,.JPEG,.JBIG2,.JB2}

\newcommand{\Sigmoidapp}{\ensuremath{\operatorname{\widehat{Sig}}}\xspace}
\newcommand{\Sigmoid}{\ensuremath{\operatorname{Sig}}\xspace}
\newcommand{\softmax}{\ensuremath{\operatorname{softmax}}\xspace}
\newcommand{\conf}{\ensuremath{\operatorname{conf}}\xspace}

\newcommand{\class}{\ensuremath{\operatorname{class}}\xspace}
\newcommand{\ReLU}{\ensuremath{\operatorname{ReLU}}\xspace}
\newcommand{\inp}{\ensuremath{\operatorname{in}}\xspace}
\newcommand{\out}{\ensuremath{\operatorname{out}}\xspace}
\newcommand{\weight}{\ensuremath{\operatorname{weight}}\xspace}
\newcommand{\actf}{\ensuremath{a}\xspace}

\newcommand{\cond}{\ensuremath{\operatorname{cond}}\xspace}

\DeclareMathOperator*{\argmax}{arg\,max}
\DeclareMathOperator*{\LSE}{LSE}
\newcommand{\defn}{\ensuremath{\stackrel{\textup{\tiny def}}{=}}}

\begin{document}


\title{Verifying Global Two-Safety Properties in Neural Networks with Confidence}

\author{%
Anagha Athavale$^{1,2}$, Ezio Bartocci$^{1}$, Maria Christakis$^{1}$, Matteo Maffei $^{1}$, Dejan Nickovic$^{2}$ and Georg Weissenbacher$^{1}$%
}
\institute{ TU Wien, Austria \and AIT Austrian Institute of Technology, Austria}

\maketitle

\thispagestyle{plain}
\pagestyle{plain}

{\def\thefootnote{}\footnotetext{Accepted at the 36th International Conference on Computer Aided Verification, 2024.}}

\begin{abstract}
We present the first automated verification technique for  confidence-based 2-safety properties, such as global robustness and global fairness, in  deep neural networks (DNNs). Our approach combines self-composition to  leverage existing reachability analysis techniques and  a novel abstraction of the softmax function, which is amenable to automated verification. We characterize and prove the soundness of our static analysis technique. Furthermore, we  implement  it on top of Marabou, a safety analysis tool for neural networks, conducting a performance evaluation on several publicly available benchmarks for DNN verification.  

\end{abstract}


\section{Introduction}
\label{sec:intro}

Deep neural networks (DNNs)
\cite{gurney1997introduction,goodfellow2016deep} 
 encountered tremendous success in the recent past due to their ability to infer highly nonlinear relations from data, learn accurate predictive models, and make smart decisions with little or no human intervention. Despite their success, the correctness of neural networks remains a major concern due to their complexity and lack of transparency. 
This is especially the case for safety- and security-critical applications where errors and biases can have serious undesired consequences, such as in medical diagnosis \cite{amato2013artificial}, self-driving cars \cite{rao2018deep}, or financial systems \cite{duan2019financial}.  New techniques that can prove mathematical guarantees concerning the behavior of neural networks are the need of the hour\cite{TranXJ22}. 
An effective approach to address this issue is the use of automatic verification techniques, which can either formally prove that the network adheres to a specified property or return a concrete input (witness) demonstrating a violation of the property\cite{bjesse2005formal}. 

\emph{Robustness} and \emph{fairness} are two important properties of neural networks. Robustness refers to the neural network's ability to make accurate predictions even in the presence of input perturbations. In particular, a robust neural network is able to produce accurate results without being overly sensitive to small changes in the input.
Fairness, on the other hand, refers to a neural network's ability to make unbiased and equitable predictions, particularly in cases where the input data may contain sensitive attributes such as gender, race, or age. A neural network that is not fair may produce biased results that discriminate against certain groups, which can have serious ethical and social implications.

\paragraph{Local robustness and fairness} provide the dominant perspective in  verification and adversarial testing of DNNs. Local robustness \cite{seshia2018formal,katz2017reluplex,huang2017safety,gopinath2018deepsafe} intuitively requires the DNN to have the following property with respect to a given  input $x$ -- it has to make the same prediction for the input $x$ as for all the points in the vicinity of $x$. Local fairness \cite{urban2020perfectly,xie2023deepgemini,seshia2018formal} is defined in a similar way, with the distance metric used for the inputs being the main difference. Both properties can be formalized as safety properties. This has led to the design of a variety of SMT-based techniques \cite{pulina2012challenging,huang2017safety,li2019analyzing}, which encode the neural networks and the property to be verified as an SMT solving problem in order to enable automated verification. Other works approach the verification problem using static analysis \cite{pulina2010abstraction, singh2019abstract,gehr2018ai2,10097028} which over-approximates DNN executions, thereby compromising precision for higher scalability. Alternative verification techniques  include mixed-integer programming \cite{cheng2017maximum, tjeng2017evaluating,dutta2018output} and modified simplex algorithms~\cite{katz2017reluplex,katz2019marabou}.

\begin{figure}[h!]
  \centering
  \includegraphics[width=0.6\linewidth]{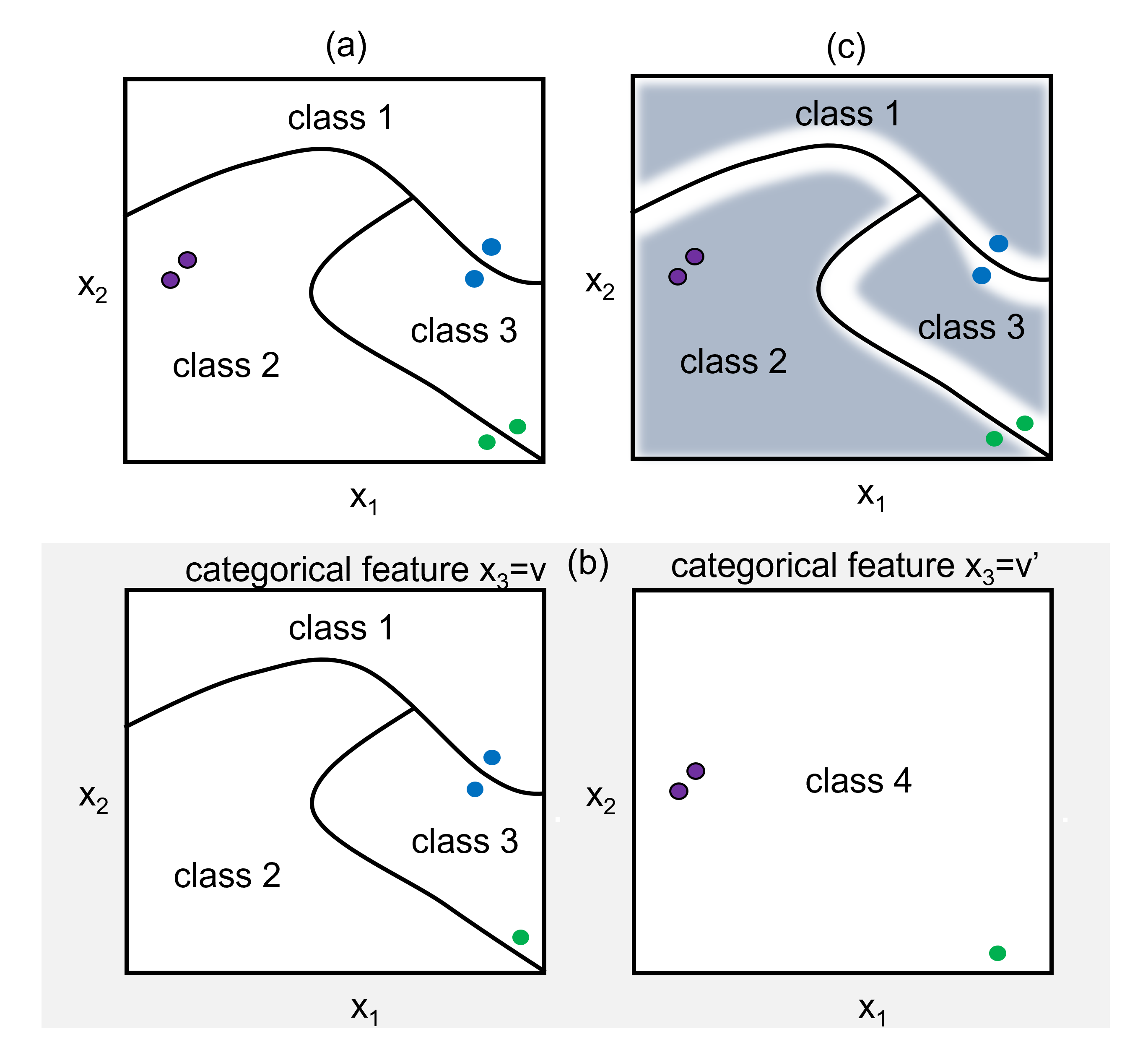}
  \caption{(a) Local, (b) partitioned-global and (c) our confidence-based robustness. $x_1$ and $x_2$ denote continuous input points, while $x_3$ denotes a categorical input in the partitioned-global approach (b). The shades of gray in (c) depict the level of confidence of the neural network with respect to the given inputs -- dark gray denotes high while white denotes low confidence level. The neural network is robust to the pair of purple points in all three cases (a), (b) and (c). The neural network is not robust for the pair of blue points in the case of local and partitioned-global (b) robustness, but is robust according to our definition (c). Finally, the neural network is not robust for the pair of green points according to both the local and our confidence-based global robustness (a) and (c), but is robust with respect to the partitioned-global robustness (b). The global partitioning method does not catch the counterexample, because the two green points are in separate partitions.}
  \label{fig:overview}
\end{figure}

\autoref{fig:overview}~(a) illustrates the properties of local robustness. It shows the classification of an input $\vec{x}$ that includes two (continuous) features $x_1$ and $x_2$. The pair of purple points is not a counterexample to robustness, as both inputs lie within the same class. The green and blue points, however, represent counterexamples to local robustness, as they fall on different sides of the decision boundaries. 

The above example shows two major limitations of local properties. First, there are always inputs arbitrarily close to the decision boundary, which then constitute  counterexamples to local robustness. Second, local robustness is defined only for a specific input. Consequently, it does not provide any guarantees for any other input. It follows that the robustness of the entire neural network cannot be assessed with local robustness only.

\paragraph{Global robustness and fairness} The limitation of the local definition for robustness and fairness indicates the need for a global property that evaluates the expected input/output relation over \emph{all pairs of inputs}. 

We first observe that global robustness and fairness of DNNs are \emph{hyperproperties}, i.e. properties that relate multiple executions of the model. 
Khedr et al. ~\cite{khedr2022certifair} and Biswas et al. ~\cite{BiswasR23} recently introduced the first verification techniques for hyperproperties in DNNs. These works assume that the inputs contain categorical variables. Based on this strong assumption, these two approaches partition the input space based on categorical features to avoid comparing inputs close to decision boundaries, which would lead to a non-satisfiable property. This is illustrated in \autoref{fig:overview}~(b). Here, we assume that $\vec{x}$ includes a categorical feature $x_3$ in addition to the continuous features $x_1$ and $x_2$. The left (right) part of \autoref{fig:overview}~(b) depicts classes and inputs in the partition based on the categorical feature $x_3$ with value $v$ ($v'$). Consequently, only pairs of inputs belonging to the \emph{same} partition are compared. Inputs belonging to two different partitions (e.g. green points in \autoref{fig:overview}~(b)) deviate in at least one categorical feature and can hence be assumed to violate the premise that these inputs are ``close''. According to this approach, a classification in a secure network can only change with different categorical values. Any two points that lie in the same partition but belong to different classes (e.g. the pair of blue points in \autoref{fig:overview}~(b)) are considered counterexamples to the global property. This leads to a strong limitation that does not admit two classes to result from continuous inputs only, as typically required for robustness. As a result, the two approaches~\cite{khedr2022certifair,BiswasR23} address only verification of global fairness.

\paragraph{Our contributions} Inspired by the work of Chen et al.~\cite{chen2021learning} on properties of rule-based security classifiers, we adopt a \emph{confidence-based} view on global robustness and fairness for DNN. The idea is to compare all input pairs which are (1) sufficiently close and (2) for which at least one of them yields a high confidence classification. This intuitive definition expects robust and fair DNNs to generate outputs with low confidence near the decision boundary. 

We therefore propose \emph{confidence-based 2-safety property}, the first definition that unifies global robustness and fairness for DNNs. Our definition highlights the hyperproperty nature of global properties and uses the confidence in the DNN as a first class citizen.

We briefly illustrate the intuition behind our confidence-based 2-safety property definition with focus on robustness in \autoref{fig:overview}~(c), in which the input space is colored into shades of gray and where every gray value corresponds to a confidence of the network. Darker shades of gray represent higher levels of confidence for the given classification. Our definition captures two reasonable assumptions: (1) continuous inputs can also trigger changes in classification, and (2) the confidence of the neural network at a decision boundary must be relatively low. In essence, our definition requires that for any input with high-confidence, all its $\epsilon$-neighbour inputs yield the same class (e.g. the two purple points in \autoref{fig:overview}~(c)). This notion discards inputs near the decision boundaries as counterexamples, as long as they result in outputs with low confidence (e.g. the two blue points in \autoref{fig:overview}~(c)).  Systems satisfying the 2-safety properties hence guarantee that input points classified with a high confidence are immune to adversarial perturbation attacks. In \autoref{fig:overview}~(c), the pair of green inputs witness the violation of the confidence-based 2-safety property -- the two points lie in different classes and one of them has an output with high confidence.

This confidence-based view makes a conceptual change to the definition of global properties, as it requires relating not only inputs, but also confidence values to the outputs. This conceptual change poses a significant challenge to the verification problem because checking a confidence-based property on a DNN requires reasoning about its \emph{softmax} layer, which is not supported by the state-of-the-art DNN verification tools~\cite{wang2021beta, xu2020fast, katz2019marabou, zhang2018efficient, ferrari2022complete, bak2020improved}. To solve this problem, we develop the first verification method that supports DNNs with softmax, in which we use a linearized over-approximation of the softmax function. We then combine it with self-composition~\cite{BartheDR11} in order to verify confidence-based 2-safety properties. We formally prove the soundness of our analysis technique, characterizing, in particular, the error bounds of our softmax over-approximation. 

We demonstrate our approach \footnote{ \url{https://github.com/anaghaathavale/Global_2Safety_with_Confidence.git}}in Marabou~\cite{katz2019marabou}, a state-of-the-art analysis tool for local robustness based on a modified simplex algorithm, which we extend to support global robustness and global fairness. We show that by combining our method with binary search, we can go beyond verification and synthesize the minimum confidence for which the DNN is globally robust or fair. We finally conduct a performance evaluation on four neural networks trained with publicly available datasets to demonstrate the effectiveness of our approach in identifying counterexamples and proving global robustness and fairness properties.

\section{Background}
\label{background_section}

\subsection{Feed-Forward Neural Networks}

In feed-forward neural networks data flows uni-directionally, which means there are no back edges.  An input layer receives the inputs that move via one or multiple hidden layers to the output layer~\cite{huang2003learning}. A layer has multiple neurons, each connected to the neurons in the next layer using a set of weights. Each layer also has an associated bias. Weight and bias selection is crucial to the performance of a neural network and is performed during the training phase. Outputs are calculated by processing the inputs using weights and biases, followed by applying the activation functions and then propagating the processed inputs through the network\cite{sharma2017activation}. 

Formally, a feed-forward neural network $f: \mathbb{R}^m \to \mathbb{R}^n$ is modeled as a directed acyclic graph $G=(V,E)$ that consists of a set (finite) of nodes $V$ and a set of edges $E\subseteq V\times V$.%
\footnote{We follow the formalization of \cite{albarghouthi2021introduction} and omit the biases for simplicity.}%
The nodes $V$ are partitioned into $l$ layers $V^i$ with $1\leq i\leq l$, where $V^1$ and $V^l$ represent the input and output layers, and $V^2,\ldots, V^{l-1}$ represent the hidden layers, respectively. We use $v_{i,j}$ to denote node $j$ in layer $i$. The edges $E$ connect nodes in $V^{i-1}$ with their successor nodes in $V^i$ (for $1< i \leq l$).  

Each node $v_{i,j}$ has an input and an output, where the latter is derived from the former by means of an activation function. We use $\inp(v_{i,j})$ and $\out(v_{i,j})$ to denote the input and output value of node $v_{i,j}$, respectively. The output is determined by
\begin{equation}
\label{eq:activation_function}
    \out(v_{i,j}) = \actf_{i,j}(\inp(v_{i,j}))
    )\,,
\end{equation}
where $\actf_{i,j}$ is the activation function. The input to node $v_{i,j}$ in layer $V^i$ is determined by the outputs of its predecessors $v_{i-1,1},\ldots,v_{i-1,k}$ in $V^{i-1}$ and weights associated with the edges $(v_{i-1,k},v_{i,j})\in E$ for $1\leq k\leq \vert V^{i-1}\vert$:

\begin{displaymath}
    \inp(v_{i,j})=\sum_{k=1}^{\vert V^{i-1}\vert} \weight((v_{i-1,k},v_{i,j}))
    \cdot \out(v_{i-1,k})
\end{displaymath}

The values of the nodes in the input layer $V^1$ are determined by the input $\vec{x}$ to $f(\vec{x})$, i.e.,  
\begin{displaymath}
(\inp(v_{1,1}),\ldots,\inp(v_{1,m}))=\vec{x}\,.
\end{displaymath}

The output of the final layer $V^l$ is then computed by propagating the inputs according to the activation functions (see \autoref{eq:activation_function} above). Consequently, a graph $G$ with $\vert V^1\vert=m$ input and $\vert V^l\vert=n$ output nodes induces a function $f:\mathbb{R}^m \to \mathbb{R}^n$ whose semantics is determined by the activation functions. 

In this paper, we concentrate on the Rectified Linear Unit (\ReLU) activation function, which is frequently applied to the hidden layers of deep neural networks. For a (scalar) input value $x$, \ReLU returns the maximum of 0 and $x$, i.e.
\begin{displaymath}
    \ReLU(x)= \max(0, x)\,.
\end{displaymath}

In neural networks that are used as classifiers and map an input $\vec x$ to one of $m$ labels in a set of classes $C$, the final layer typically employs a \softmax function to ensure that the output represents normalized probabilities corresponding to each of the $n$ classes. Mathematically, 
\begin{equation} \label{eq:softmax} 
    \softmax(\vec{z})_i = \frac{e^{z \textsubscript i}}{\sum_{j=1}^{n} e^{z\textsubscript j}}
\end{equation} 
where $\vec{z}$ represents the values $\out(v_{l-1,i})$ for $1 \leq i \leq n$ and $n=\vert V^{l-1}\vert$, and $z_i$ is the $i^{\text{th}}$ element in $\vec{z}$. This induces a function 
$y: \mathbb{R}^n \to [0,1]^n$ mapping every output of $V^{l-1}$ to a confidence score in the range $[0,1]$. 
Consequently, $f(\vec{x})$ outputs a probability distribution over the possible labels in $C$, where each component of the output vector represents the probability of input $\vec{x}$ belonging to the corresponding class. We use $\conf(f(\vec x))$ to refer to the highest probability value in the \softmax layer of $f(\vec x)$ and call it the {\em confidence}, i.e.,
\begin{equation}
    \label{eq:def_confidence}
    \conf(f(\vec{x}))=\max(\out(v_{l,1}),\ldots,\out(v_{l,n}))
\end{equation}
Finally, a function $\class: \mathbb{R}^m \to C$ then maps the output of $f$ to the class $C$ corresponding to the highest probability in $f(\vec x)$:
\begin{equation}
    \class(f(\vec{x})) = \argmax_{1\leq i\leq n}(\out(v_{l,i}))
\end{equation}

\subsection{Hyperproperties} 

Hyperproperties \cite{clarkson2010hyperproperties} are a class of properties that capture relationships between multiple execution traces. This is in contrast to traditional properties, which are evaluated over individual traces. 

To define traces in the context of feed-forward neural networks, we extend our notation $\out$ to layers as follows:
\vspace{-0.5ex}
\begin{displaymath}
    \out(V^i)=(\out(v_{i,1}),\ldots,\out(v_{i,k}))
\end{displaymath}
where $k=\vert V^i\vert$. Let $\inp(V^i)$ be defined similarly. The corresponding trace $\pi$ for $f(\vec x)$ is then formally defined as
\begin{displaymath}
\pi = \inp(V^1),\out(V^1),\ldots,\inp(V^l),\out(V^l)
\end{displaymath}
where $\inp(V^1)=\vec{x}$.

Note that each execution is entirely determined by the input value $\vec x$ (assuming that the function $f$ implemented by the network is deterministic). Quantifying over traces $\pi$ of $f(\vec{x})$ hence corresponds to quantifying over the corresponding inputs $\vec{x}$.
A traditional safety property would then quantify over the inputs $\vec{x}$, e.g.
 \begin{displaymath}
\forall \vec x \,.\,\conf(f(\vec x))\geq \kappa\,,
\end{displaymath}
stating that the confidence of each classification of the network should be larger than a threshold $\kappa$.
Another example of a traditional safety property is local robustness, given in Definition \ref{local_rob} in \autoref{robustness_sec}.

A hyperproperty, on the other hand, refers to, and quantifies over,  more than one trace. An example would be
\begin{equation}
\label{lipschitz_eq}
\forall \vec{x},\vec{x}'\,.\,
\frac{\vert f(\vec{x})_i-f(\vec{x}')_i\vert}{||\vec{x}-\vec{x}'||}\leq K_i\,, 1\leq i \leq n
\end{equation}
where $f(\vec{x})_i$ denotes $\out(v_{l,i})$. \autoref{lipschitz_eq} states that $K_i$ sets the maximum limit of the Lipschitz constant for $f(\vec{x})_i$. A hyperproperty central to this paper is global robustness, defined in Definition \ref{global_rob} \autoref{robustness_sec}.

Hyperproperties are used to capture important properties that involve multiple inputs, such as robustness and fairness. By verifying hyperproperties of neural networks, we can ensure that they behave correctly across all possible input traces.

\subsection{Relational Verification and Self-Composition} 
\label{self_composition}

Hyperproperties are verified by means of so-called relational verification techniques: the idea is to verify if $k$  program executions  satisfy a given property \cite{barthe2011relational}, expressing invariants on  inputs and outputs of such executions. 
Several  security properties (e.g., information flow) can be expressed by relating two  executions of the same program differing in the inputs: such properties are called 2-safety properties. Global robustness in neural networks can also be seen as a 2-safety property \cite{seshia2018formal}.

2-safety properties can be verified in a generic way by self-composition \cite{barthe2011relational}: the idea is to compose the program with itself and to relate the two executions. 
In the context of neural networks, the self-composition of a network $f$ is readily defined as a function over 
\begin{equation}    
    f(\vec{x})\times f(\vec{x}') = \lambda (\vec{x},\vec{x}')\,.\,(f(\vec{x}),f(\vec{x}'))
\end{equation}
where $(\vec{x},\vec{x}')$ denotes the concatenation of the vectors $\vec{x}$ and $\vec{x'}$ and $\lambda \vec{x}\,.\,f(\vec x)$ denotes the lambda term that binds $\vec{x}$ in $f(\vec{x})$. The underlying graph $G=(V,E)$ is simply duplicated, i.e., we obtain a graph $G\times G'=(V\cup V', E\cup E')$ where $V'$ and $E'$ are primed copies of $V$ and $E$.

A counterexample to a universal 2-safety property over the self-composition of $f$ comprises of a pair of traces of $f$ witnessing the property violation.

\subsection{Robustness and Fairness}
\label{robustness_sec}

Robustness in neural networks refers to the ability of a model to perform consistently in the presence of small perturbations of the input data. The common approach to address robustness in neural networks is to define it as a \emph{local robustness} \cite{bastani2016measuring} property. 
 For an input $\vec x$,
 a neural network is locally robust if it yields the same classification for $\vec x$ and all inputs $\vec x'$ within distance $\epsilon$ from $\vec x$ \cite{mangal2019robustness}:

\begin{definition}[Local Robustness\label{local_rob}]
A model $f$ is locally $\epsilon$-robust at point $\vec x$ if 
$$\forall \vec{x}'\,.\, ||\vec{x}-\vec{x}'|| \leq \epsilon \to \class(f(\vec{x})) = \class(f(\vec{x}'))$$
\end{definition} 

Local robustness, therefore, is defined only for inputs within a distance $\epsilon$ of a specific $\vec x$ and, thus, does not provide global guarantees. Here  $||\cdot||$  represents the  distance metric used over the input  space.
Intuitively, global robustness tackles this problem by requiring that the local robustness property must hold for {\em every} input within the input space \cite{seshia2018formal}. 
Definition \ref{global_rob} gives the general definition of global robustness used in \cite{leino2021globally, chen2021learning}. It essentially states that all input points in a small neighborhood $\epsilon$, are mapped to the same class.

\begin{definition}[Global robustness]\label{global_rob}
A model $f$ is globally $\epsilon$-robust if 
\begin{displaymath}
    \forall \vec x, \vec x'\,.\, ||\vec x - \vec x'|| \leq \epsilon \to \class(f(\vec x)) = \class(f(\vec x'))\,
\end{displaymath}
\end{definition}

Clearly, global robustness as formalized in Definition \ref{global_rob} makes sense only for selected distance metrics, which in particular avoid comparing inputs close to the decision borders. For instance, \cite{leino2021globally} addresses this by introducing an additional class $\bot$ to which $\class(f(\vec x))$ evaluates whenever the difference between the highest and second-highest probability falls below a certain threshold (determined by the Lipschitz constants of $f$). The global robustness requirement is then relaxed at these points. 

\begin{definition}[Global fairness]\label{global_fair} A model is said to be globally fair if:
\begin{multline*}
    \forall \vec{x} = (x_s, \vec{x_n}), \vec{x'} = (x_s', \vec{x_n'}). \, \\ ||\vec{x_n} – \vec{x’_n}|| \leq \epsilon \; \wedge (x_s \neq x_s') \to \class(f(\vec x)) = \class(f(\vec x'))
\end{multline*}
 where $x_s$ and $x_n$ are the sensitive and non-sensitive attributes of $\vec{x}$, respectively.
\end{definition}
\cite{khedr2022certifair,BiswasR23} address a similar problem, which arises in the context of fairness, by partitioning the input space based on categorical features. 
In general, if the input to a decision-making neural network comprises of certain sensitive attributes, say age or gender, the network is said to be fair if the sensitive attributes do not influence its decisions \cite{seshia2018formal}. Definition \ref{global_fair} gives the general definition of global fairness used in \cite{khedr2022certifair,BiswasR23}. 


Ensuring fairness in neural networks is important because these models are increasingly being used in decision-making processes that can have significant impacts on peoples' lives. For example, a hiring algorithm that discriminates against certain groups of job applicants based on their race or gender could perpetuate existing biases and inequalities in the workplace \cite{binns2018fairness}.



\section{Confidence Based Global Verification of Feed-Forward Neural Networks}
\label{sec:technical_content}


We now formalize confidence-based 2-safety property, the first definition
that unifies global robustness and fairness for DNNs in Definition \ref{conf_rob}. It is a hyperproperty that takes the confidence of the decision into account when checking for the safety of the network. Before we give the actual definition, we introduce additional notation. Given an input $\vec{x} = (x_1, \ldots, x_n)$, we assume that its every component $x_i$ is either a \emph{categorical} or \emph{real} value. We then define the distance $d(x_i, x'_i)$ as $|x_i - x'_i|$ when $x_i$ is real-valued. We use instead the following distance: 
$$
d(x_i, x'_i) = \begin{cases}
      0, & \text{if}\ x_i = x'_i \\
      1, & \text{otherwise}
    \end{cases}
$$
\noindent when $x_i$ is a categorical value. We define $\emph{cond}(\vec{x}, \vec{x}', \vec{\epsilon})$ as a (generic) Boolean \emph{condition} that relates inputs $\vec{x}$ and $\vec{x}'$ to a tolerance vector $\vec{\epsilon}$. 

\begin{definition}[Confidence-based global 2-safety]\label{conf_rob}
A model $f$ is said to be globally 2-safe for confidence $\kappa > 0$ and tolerance $\vec{\epsilon}$ iff 

\begin{displaymath}
\begin{aligned}
\forall \vec{x}, \vec{x}'\,.\,
      \cond(\vec{x}, \vec{x}', \vec{\epsilon})\; & \wedge \conf(f(\vec{x})) > \kappa & \implies \class(f(\vec{x})) = \class(f(\vec{x}'))
      \end{aligned}
\end{displaymath}
\end{definition}

Next, we instantiate the above 2-safety property for confidence-based global robustness and fairness.

For \emph{confidence-based global robustness}, \cond is defined as: 

$$\cond(\vec{x}, \vec{x}', \vec{\epsilon}) = \bigwedge_{ i \in [1,n]}  d(x_i, x’_i) \leq \epsilon_i$$

For \emph{confidence-based global fairness}, $\vec{x}$ can be split into sensitive $\vec{x_s}$ and non-sensitive $\vec{x_n}$ attributes. For confidence-based global fairness,
\begin{displaymath}
\begin{aligned}
\cond(\vec{x}, \vec{x}', \vec{\epsilon}) =  \bigwedge_{x_i \in \vec{x_s}} d(x_i, x’_i) > 0 \; \wedge \bigwedge_{x_i \in \vec{x_n}} d(x_i, x'_i) \leq \epsilon_i
\end{aligned}
\end{displaymath}

\noindent where for any categorical $x_i \in \vec{x_n}$, its associated tolerance threshold $\epsilon_i=0.5$.

Intuitively, confidence-based global fairness ensures that for any data instance, $x$ classified with high confidence $\kappa$, no other data instance, $x'$, that only differs with $x$ in the value of the sensitive attribute (e.g. age, gender, ethnicity) shall be classified to a different class. 

As defined in Section \ref{background_section}, $f(x)$ represents the feed-forward neural network, which maps inputs to classes with corresponding confidence scores. By introducing the threshold $\kappa$, our definition effectively ignores classification mismatches that arise from decisions with low confidence. The rationale is as follows:
\vspace{-1ex}
\begin{itemize}
\item Different classifications close to decision boundaries need to be allowed, as safety can otherwise only be satisfied by degenerate neural networks that map all inputs to a single label.
\item On the other hand, input points classified with a high confidence should be immune to adversarial perturbations and also uphold fairness.
\end{itemize}

\subsection{Encoding 2-Safety Properties as Product Neural Network}

In this section, we reduce checking of the 2-safety hyperproperty in Definition \ref{conf_rob} to a safety property over a single trace.
Given a neural network $f$ (as defined earlier in section \ref{background_section}), the product neural network is formed by composing a copy of the original neural network with itself. Checking 2-safety then reduces to checking an ordinary safety property for the self-composed neural network that consists of two copies of the original neural network, each with its own copy of the variables. 

The product neural network is now treated as the model to be verified. A product network allows the reduction of a hyperproperty to a trace property, thereby reducing the problem of hyperproperty verification to a standard verification problem, which can be solved using an existing standard verification technique.

\begin{table*}[t]\centering
\begin{tabular}{lll}
$\mathsf{Equation}$ & ::= & $\mathsf{Sum}\,\Diamond\, \mathsf{Constant} $ with $\Diamond\in\{\leq, \geq, =\}$\\
&& and $\mathsf{Sum} ::= \mathsf{Sum} + \mathsf{Constant}\cdot\mathsf{Variable}~\vert~\mathsf{Constant} \cdot \mathsf{Variable}$ \\
$\mathsf{MaxConstraint}$ & ::= & $\mathsf{Variable} = \max(\mathsf{VariableList})$ \\
&& and $\mathsf{VariableList} ::= \mathsf{VariableList},  \mathsf{Variable}~\vert~\mathsf{Variable} $\\
$\mathsf{AbsConstraint}$ & ::= & $\mathsf{Variable} = \vert\mathsf{Variable}\vert$  \\
$\mathsf{ReLUConstraint}$ & ::= & $ \mathsf{Variable} = \ReLU(\mathsf{Variable})$  \\
$\mathsf{Disjunction}$ & ::= & $(\mathsf{Disjunction}\vee\mathsf{Equation})~\vert~\mathsf{Equation}$
\end{tabular}
\caption{Marabou's piecewise linear constraints\label{marabou_grammar}}
\end{table*}
\vspace{-1ex}
\paragraph{Product Neural Network}

We encode $f(\vec{x})$ using piecewise linear constraints (see \autoref{marabou_grammar}). Each node $v_{i,j}$ is represented by two variables $\mathsf{in}_{i,j}$ and $\mathsf{out}_{i,j}$ representing its input and output, respectively. Inputs and outputs are related by the following constraints:
\begin{displaymath}
\mathsf{in}_{i,j}=\sum_{k=1}^{\vert V^{i-1}\vert} w_{i,j}^{i-1,k} \cdot\mathsf{out}_{(i-1,)k}\quad \wedge\quad \mathsf{out}_{i,j} = a_{i,j}(\mathsf{in}_{i,j})
\end{displaymath}
where $w_{i,j}^{i-1,k}$ is the weight associated with the edge $(v_{i-1,k},v_{i,j})$ and $a_{i,j}$ is the activation function of node $v_{i,j}$.
To encode the self-composition, we duplicate all variables and constraints by introducing primed counterparts $\mathsf{in}'_{i,j}$ and $\mathsf{out}'_{i,j}$ for $\mathsf{in}_{i,j}$ and $\mathsf{out}_{i,j}$.
\vspace{-2ex}
\paragraph{Transfer Functions and Operators} \label{transfer_fun_and_op}

\ReLU{}s can be readily encoded using $\mathsf{out}_{i,j}=\ReLU(\mathsf{in}_{i,j})$. There is, however, no direct way to encode \softmax using the constraints in \autoref{marabou_grammar}, hence we defer the discussion to \autoref{softmax_sec}. 

The $\conf$ operator can be implemented using the $\max$ constraint (cf. \autoref{eq:def_confidence}). The operator $\class$ as well as the implication, on the other hand, are not necessarily supported by  state-of-the-art static analysis tools for DNNs. For instance, they are not supported by Marabou~\cite{katz2019marabou}, on which we base our implementation. For reference, \autoref{marabou_grammar} illustrates the linear constraints supported by Marabou. We thus introduce an encoding, which we detail below. 

First, checking the validity of the implication in \autoref{conf_rob} can be reduced to checking the unsatisfiability of  
\begin{equation}
     \cond(\vec{x}, \vec{x'}, \vec{\epsilon}) \wedge \conf(f(\vec{x})) > \kappa  \wedge \class(f(\vec{x})) \neq \class(f(\vec{x}'))
\end{equation}

However, the grammar in \autoref{marabou_grammar} provides no means to encode disequality or $\class$ (which returns the {\em index} of the largest element of a vector). To implement disequality, we perform a case split over all $n=\vert V^l\vert$ labels by instantiating  the encoding of the entire network over $\mathsf{out}_{l,i}$ and $\mathsf{out}'_{l,i}$ for $1\leq i\leq n$. To implement this in Marabou, we execute a separate query for every case.

To handle the operator \class, we can encode the disequality $\class(f(\vec{x})) \neq \class(f(\vec{x}')$ as: 
\begin{multline}
      \ldots\;\wedge\;\overbrace{\max (\mathsf{out}_{l,1},\ldots,\mathsf{out}_{l,n})}^{\conf(f(\vec{x}))} > \kappa\; \wedge   
      \left(\max (\mathsf{out}_{l,1},\ldots,\mathsf{out}_{l,n})-\mathsf{out}_{l,i} = 0\right)\; \wedge \\  
    \left(\max (\mathsf{out'}_{l,1},\ldots,\mathsf{out'}_{l,n})-\mathsf{out'}_{l,i} \neq 0\right)\;
\end{multline}

The constraint $(\max (\mathsf{out}_{l,1},\ldots,\mathsf{out}_{l,n})-\mathsf{out}_{l,i}=0)$ ensures that $\out_{l,i}$ corresponds to the largest element in $f(\vec{x})$ (and hence that $\class(f(\vec{x})=i$). Consequently, if $(\max (\mathsf{out'}_{l,1},\ldots,\mathsf{out'}_{l,n})-\mathsf{out'}_{l,i} \neq 0)$, then we can conclude that $\class(f(\vec{x}'))\neq i$ and hence the safety constraint is violated. 

Since, Marabou does not support the disequality operator, we check whether $(\max(\mathsf{out'}_{l,1},\ldots,\mathsf{out'}_{l,n})-\mathsf{out'}_{l,i} < 0 ) $ and $(\max(\mathsf{out'}_{l,1},\ldots,\mathsf{out'}_{l,n})-\mathsf{out'}_{l,i} > 0 ) $, and if both constraints are not satisfied, we know that $(\max(\mathsf{out'}_{l,1},\ldots,\mathsf{out'}_{l,n})-\mathsf{out'}_{l,i} \neq 0 )$. 

While the above transformation is equivalence preserving, the encoding of \softmax requires an approximation, described in the following subsection.

\subsection{Softmax Approximation}
\label{softmax_sec}
\subsubsection{Softmax in terms of $\max$ and $\Sigmoid$}
We can approximate softmax using a $\max$ operator and a sigmoid function as follows.
Consider ${\softmax (\vec{z})\textsubscript i}$ (cf.~Equation \ref{eq:softmax}), for i=1,
\begin{align} \label{sigma_for_i=1}
{\softmax (\vec{z})\textsubscript 1}  &= \frac{1}{1+ (e^{z{\textsubscript{\tiny 2}}} + \dotsb + e^{z{\textsubscript{\tiny n}}})e^{-z{\textsubscript{\tiny 1}}} } \\&= \frac{1}{1+ e^{log (e^{z{\textsubscript{\tiny 2}}} + \dotsb + e^{z{\textsubscript{\tiny n}}})}e^{-z{\textsubscript{\tiny 1}}}} = \frac{1}{1+ e^{(-z{\textsubscript{\tiny 1}} + log (e^{z{\textsubscript{\tiny 2}}} + \dotsb + e^{z{\textsubscript{\tiny n}}}))}}
\end{align}

We can now generalize the result from (\ref{sigma_for_i=1}) for i
\begin{equation} \label{sig_as_logsum}
\begin{aligned}
&\softmax(\vec{z})\textsubscript i = \frac{1}{1+ e^{(-z{\textsubscript{\tiny i}} + log (\sum_{j\neq i}^{n} e^{z{\textsubscript{\tiny j}}}))}} 
=\Sigmoid(z{\textsubscript{\tiny i}} -  \mathop{\LSE_{1}^{n}}_{j\neq i}(z_j))
\end{aligned}
\end{equation}

\noindent where LSE (the \emph{log-sum-exp}) is:
 $$\mathop{\LSE_{1}^{n}}_{j\neq i}(z_j) = log (\sum_{j=1, j\neq i}^{n} e^z\textsubscript{\tiny{j}}) \mbox{ and } \Sigmoid(x) = \frac{1}{1+ e^{-x}}$$ 

We know from \cite{pant2017smooth} that LSE is bounded:
\begin{equation}
\begin{split}
\mathop{max_{1}^{n}} (z_i) \leq \mathop{\LSE_{1}^{n}} (z_i)\leq \mathop{max_{1}^{n}} (z_i) +log(n)
\end{split}
\end{equation}
with $\mathop{max_{1}^{n}} (z_i)=max(z_1, \cdots, z_n)$, in particlular, when $z_1=\cdots=z_n$, we have: 
\begin{equation}
\begin{split}
\mathop{\LSE_{1}^{n}} (z_i) = \mathop{max_{1}^{n}} (z_i) +log(n)
\end{split}
\end{equation}
Then softmax has as lower bound:
\begin{equation}
\begin{split}
 \softmax (\vec{z})\textsubscript i \geq \Sigmoid(z{\textsubscript{\tiny i}} -  \mathop{max_{1}^{n}}_{j\neq i}(z_j) + log(n-1))
\end{split}
\end{equation}
and as upper bound:
\begin{equation}
\begin{split}
 \softmax (\vec{z})\textsubscript i \leq \Sigmoid(z{\textsubscript{\tiny i}} -  \mathop{max_{1}^{n}}_{j\neq i}(z_j) )
\end{split}
\end{equation}
When $n=2$, the softmax is equivalent to the sigmoid:

\begin{equation}
\begin{aligned}
 \softmax (\vec{z})\textsubscript 1 = \Sigmoid(z{\textsubscript{\tiny 1}} -  z_2) \mbox{ and }  \softmax (\vec{z})\textsubscript 2 = \Sigmoid(z{\textsubscript{\tiny 2}} -  z_1) 
\end{aligned}
\end{equation}

Now that we know how to approximate a softmax using a sigmoid and max, we need to find a piece-wise linear approximation of sigmoid since sigmoid is also a non-linear exponential function.

\subsubsection{Piece-wise approximation of sigmoid}
We approximate sigmoid as a piece-wise linear function using the Remez exchange algorithm\cite{remez}. The Remez algorithm is an iterative algorithm that finds simpler approximations to functions. It aims to minimize the maximum absolute difference between the approximated polynomial and the actual function. The algorithm takes a maximum acceptable error $\delta$ and generates $l$ linear segments to approximate the sigmoid function such that the error is less than $\delta$.
We use the Remez algorithm to approximate 
the sigmoid in the interval $[\Sigmoid^{-1}(\delta),\Sigmoid^{-1}(1-\delta)]$, where $\Sigmoid^{-1}$ is the inverse function of the sigmoid. The inverse of sigmoid is the logit function i.e., $\Sigmoid^{-1}(y)=logit(y)=log (y)/(1-y)$. For example, if the user sets $\delta$ to 0.0006, then the input domain for the algorithm lies in $[-7.423034723582278, 7.423034723582278]$.  Thus, the approximated sigmoid is: 
$$|\Sigmoidapp(x) - \Sigmoid(x)| \leq \delta$$ 
We approximate $\softmax$  with its lower bound:
\begin{equation}
\label{eq:lowerbound}
\begin{split}
 \widehat{\softmax}  (\vec{z})\textsubscript i = \Sigmoidapp(z{\textsubscript{\tiny i}} -  \mathop{max_{1}^{n}}_{j\neq i}(z_j) + log(n-1)) - \delta 
\end{split}
\end{equation}
and the upper bound for the softmax is:
\begin{equation}
\begin{split}
 \widehat{\softmax}  (\vec{z})\textsubscript i 
 \leq 
\softmax  (\vec{z})\textsubscript i \leq \Sigmoidapp(z{\textsubscript{\tiny i}} -  \mathop{max_{1}^{n}}_{j\neq i}(z_j)) + \delta 
\end{split}
\end{equation}


\begin{theorem}\label{tm}
   Let $\softmax$ and $\widehat{\softmax}$ compute the real and linearly approximated softmax (with precision $\delta$), respectively for the last layer of $n \geq 2$ neurons $\vec{z}$ of a neural network and $z_i = max(z_1, \cdots, z_n)$. Then, we have the following result: 
   \begin{equation*}
   \label{eq:conf}
   \forall \vec{z} . \ \softmax(\vec{z})\textsubscript i - \widehat{\softmax}(\vec{z}) \textsubscript i \leq \frac{n-2}{(\sqrt{n-1} + 1)^2} + 2 \delta 
   \end{equation*}
\end{theorem}

\begin{proof}
    Refer Appendix \ref{sec:appendix}
\end{proof}

\begin{theorem}\label{pm} (Class consistency)
 Let $f$ and $\hat{f}$ denote the real and the approximated (with precision $\delta$) neural networks with $n\geq 2$ outputs, respectively. Then:
 $$\conf(\hat{f}(\vec{x})) > \frac{1}{2} \implies \class(\hat{f}(\vec{x})) = \class(f(\vec{x})) $$
 \end{theorem}

\begin{proof}
    Refer Appendix \ref{sec:appendix}
\end{proof}

\subsubsection{Soundness}

For the confidence-based 2-safety property discussed before, our analysis provides a soundness guarantee. This means that whenever the analysis reports that the property specified in \ref{conf_rob} holds, then the property also holds true in the concrete execution.

\begin{theorem} (Soundness) Let $f$ and $\hat{f}$ be the original neural network and over-approximated neural network, respectively. Let $b_{n,\delta}$ be 
the error bound of the approximated softmax ($b_{n,\delta} =\frac{n-2}{(\sqrt{n-1} + 1)^2} + 2 \delta$ (see Theorem \ref{tm})). Then we have the following soundness guarantee: Whenever the approximated neural network is 2-safe for $\conf(\hat{f}(\vec{x})) > (\kappa - b_{n,\delta})$, the real neural network is 2-safe for $\conf(f(\vec{x})) > \kappa$, given $\conf(\hat{f}(\vec{x})) > \frac{1}{2}$. Formally:
\begin{multline*} 
\left(\begin{split}
\forall \vec{x}, \vec{x'}. \ \cond(\vec{x}, \vec{x'}, \vec{\epsilon}) \land {\conf}(\hat{f}(\vec{x})) > (\kappa - b_{n,\delta}) \\ \implies {\class}(\hat{f}(\vec{x})) = {\class}(\hat{f}(\vec{x'})) 
\end{split}\right) \implies \\
\left(\begin{split}\forall \vec{x}, \vec{x'}. \ \cond(\vec{x}, \vec{x'}, \vec{\epsilon}) \land \conf(\vec{f(x)}) > \kappa  \\ \implies \class(f(\vec{x})) = \class(f(\vec{x'}))\end{split}\right),
\mbox{ with } \: \conf(\hat{f}(\vec{x})) > \frac{1}{2}
\end{multline*} 

\end{theorem}

\begin{proof}
    Refer Appendix \ref{sec:appendix}
\end{proof}

\section{Implementation}
For the implementation of our technique, we use the state-of-the-art neural network verification tool Marabou \cite{katz2019marabou} as our solver. In this section, we describe how we encode the confidence-based 2-safety property in Marabou. Note that such an encoding can be expressed in a similar way for virtually any off-the-shelf neural network verifier.

\noindent \textbf{Marabou\cite{katz2019marabou}}
Marabou is a simplex-based linear programming neural network verification and analysis tool. Marabou is capable to address queries about
network’s properties, such as local robustness, by encoding them into constraint satisfaction problems. It supports fully-connected feed-forward neural networks. 
A network can be encoded as a set of 
linear constraints representing the  
weighted sum of the neurons' outputs feeding the next neuron's input, and a set of non-linear constraints defining the activation functions.
A verification query to Marabou comprises a neural network along with a property that needs to be verified. This property is defined as "linear and nonlinear constraints on the network's inputs and outputs"\cite{katz2019marabou}. 
In Marabou, network's neurons are treated as variables.  As a result, the verification problem involves identifying a variable assignment that satisfies all the constraints at the same time, or establishing that such an assignment does not exist. 
The tool uses a variant of the \emph{Simplex} algorithm at its core to make the variable assignment satisfy the linear constraints.  During the execution, 
the tool adjusts the variable assignment to either fix a linear or a non-linear constraint violation.
Although the technique implemented in Marabou is sound and complete, the tool can work only with piece-wise linear activation functions (including $\ReLU$ function and the $\max$ function) to guarantee termination.
Additionally, an essential aspect of Marabou's verification approach is deduction -- specifically, deriving more precise lower and upper bounds for each variable. The tool leverages these bounds 
to relax piece-wise linear constraints into linear ones by considering one of its segments. 

The original network g is a function of the following: input parameters, neurons, neuron connection weights, layer biases, $\ReLU$ activation functions and output classes. To make Marabou amenable for verification of 2-safety properties, we need a product neural network. This means that the execution is tracked over two copies of the original network, $g$ and $g'$ (cf.~\autoref{self_composition}). Let $X_i$ denote the set of input variables to $g$ and let $X'_i$ be a set of primed copies of the variables in $X_i$. As a result, we obtain a self-composition $g(X_i) \times g'(X_i')$ of $g$ over the input variables $X_i \cup X_i'$. 

Next, we extend the output layer with softmax function in order to extract the confidence scores with which output classes are predicted. 

\noindent \textbf{Linearized Sigmoid}\label{linearized_sigmoid}
We explain our linearized sigmoid function in this subsection. This function is used to implement an approximated piece-wise linear sigmoid function. Let the outputs of the last inner layer $l-1$ be represented by $z_i$ for $1\leq i\leq n$, where $n$ is the number of output classes. In Marabou, we first encode the linear piece-wise sigmoid function which we obtain by setting the maximum acceptable error to 0.005. This provides us with a piece-wise linear approximated sigmoid with 35 segments of the form $q_j = \{ m_j \cdot z_i + c_j, \ | \ LB \leq z_i \leq UB\}$, where $z_i$ is the variable representing the output node whose confidence we want to find. We encode each segment as an equation in and represent it using a variable $q_j$. Next, we need to select the applicable segment corresponding to the value of $z_i$. Unfortunately, Marabou does not provide a conditional construct. So, we deploy the $\min$ and $\max$ functions to emulate if-then-else.

\begin{figure}[h!]
  \centering
  \includegraphics[scale=0.4]{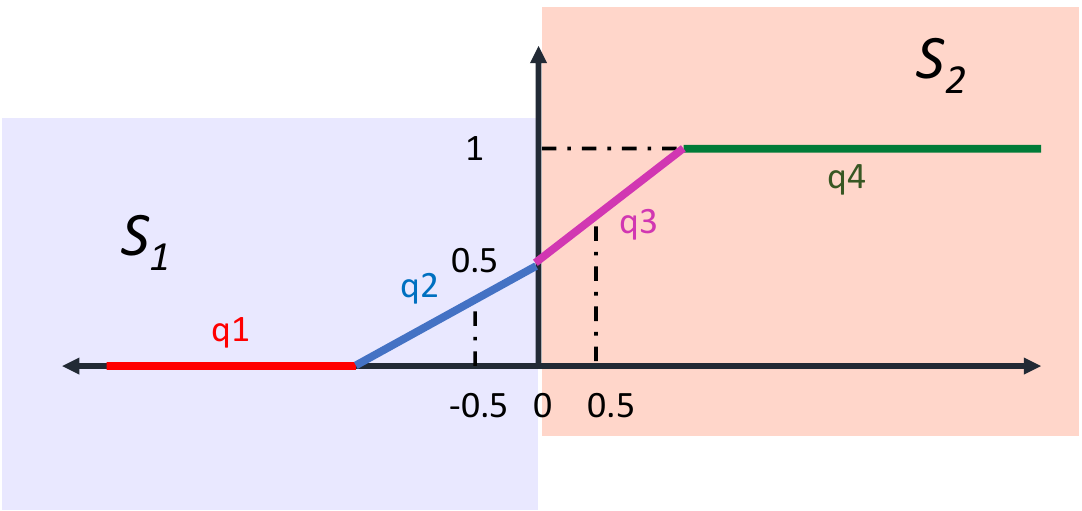}
  \caption{(Simplified) Approximation of sigmoid with 4 linear segments}
  \label{fig:min-max}
\end{figure}

First, we split the sigmoid into two convex pieces $S_1$ and $S_2$. Figure \ref{fig:min-max} illustrates this step using a simplified approximation of sigmoid with 4 linear segments $q_1$, $q_2$, $q_3$, and $q_4$. The resulting value of $S_1$ can now be expressed as $S_1 = \min(\max(0, q_1, q_2), 0.5)$. Similarly, $S_1 = \max(\min(0.5, q_3, q_4), 1)$. The values $0$ and $1$ are the minimum and maximum values of the sigmoid function and $0.5$ is the value of sigmoid at the splitting point. Second, we combine the convex segments by adding them:
$$S = \min(\max(0, q_1, q_2), 0.5)
      + \max(\min(0.5, q_3, q_4), 1) - 0.5$$
Note that we have 35 segments instead of four used in our simplified explanation.\newline

\noindent \textbf{Linearized Softmax}
The next step consists in implementing the softmax function using the output of the sigmoid function and the $\max$ function (see \autoref{eq:lowerbound}). To this end, we find the maximum of all output nodes excluding the current one, and subtract that maximum from the current output value. Finally, we apply the linearized sigmoid (cf.~\autoref{linearized_sigmoid}), to obtain the result of softmax. 

We repeat the above steps for all output nodes to obtain the softmax values corresponding to all output classes. Finally, we find the maximum value of these softmax outputs, which represents the confidence. 

\section{Experimental Evaluation}
\label{sec:eval}

For our evaluation, we used four publicly available benchmark datasets to evaluate our technique. We pre-process the datasets to remove null entries, select relevant categorical attributes, and hot-encode them. For each dataset, we train a fully connected feed-forward neural network with up to 50 neurons and ReLU activation functions. 

\textbf{German Credit:} The German Credit Risk dataset \cite{hofmann1994german} describes individuals requesting credit from a bank and classified, based on their characteristics,  in two categories ("good" or "bad") of credit risk. The dataset comprises 1000 entries.

\textbf{Adult:} The Adult dataset, also referred to as the "Census Income" dataset, is used to estimate whether a person's income surpasses \$50,000 per year based on census information \cite{Dua:2019}.

\textbf{COMPAS:} COMPAS ("Correctional Offender Management Profiling for Alternative Sanctions") is a widely-used commercial algorithm that is utilized by judges and parole officers to assess the probability of criminal defendants committing future crimes, also known as recidivism \cite{compas}.

\textbf{Law School:} The Law School Admissions Council (LSAC) provides a dataset called Law School Admissions, which includes information on approximately 27,000 law students from 1991 to 1997. The dataset tracks the students' progress through law school, graduation, and bar exams. It uses two types of academic scores (LSAT and GPA) to predict their likelihood of passing the bar exam~\cite{wightman1998lsac}.

We use TensorFlow for training neural networks and 
the NN verifier Marabou \cite{katz2019marabou} whose implementation is publicly available.  The accuracies for the deployed models are as follows:  German Credit: 0.71; COMPAS: 0.74; Law: 0.94; Adult: 0.77. In our experiments, adding more layers or nodes per layer did not result in an increased accuracy. We run all our experiments using a single AMD EPYC 7713 64-Core Processor, Ubuntu 22.04 LTS Operating System with 32 GB RAM.

\begin{figure}[!h] 
	\centering 
	\begin{minipage}[t]{4cm} 
		\centering            
		\includegraphics[scale=0.4]{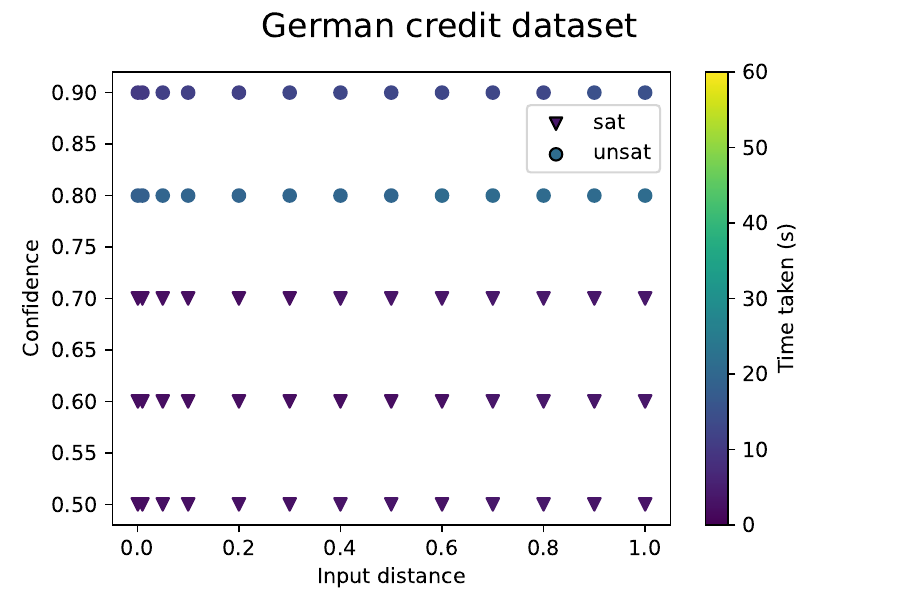} 
		\caption{Input distance vs. confidence for German credit dataset} 
            \label{fig:german_credit}
	\end{minipage} 
	\hspace{1.5cm} 
	\begin{minipage}[t]{4cm} 
		\centering 
		\includegraphics[scale=0.4]{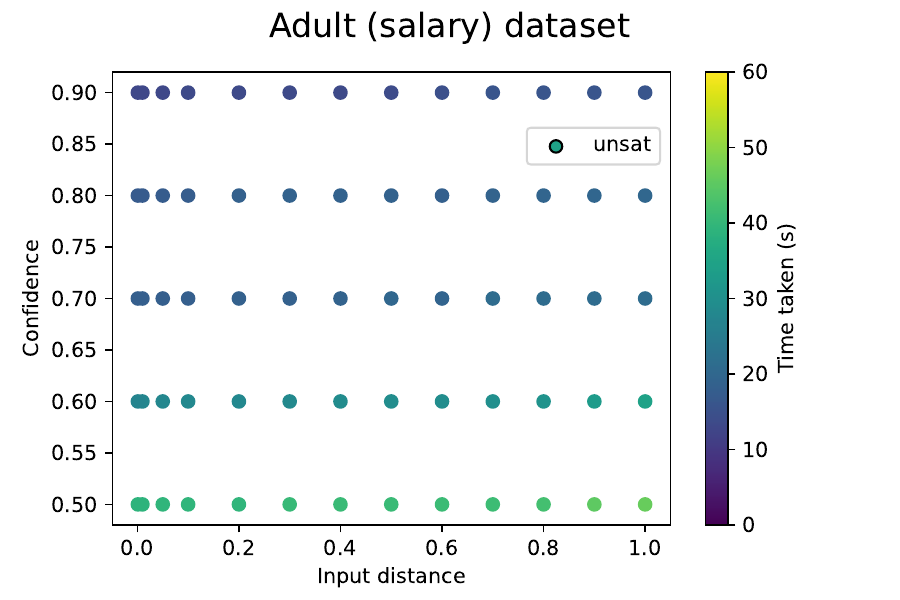} 
		\caption{Input distance vs. confidence for adult dataset} 
            \label{fig:adult}
	\end{minipage} 
        
\end{figure} 

\begin{figure}[!h] 
	\centering 
	\begin{minipage}[t]{4cm} 
		\centering            
		\includegraphics[scale=0.4]{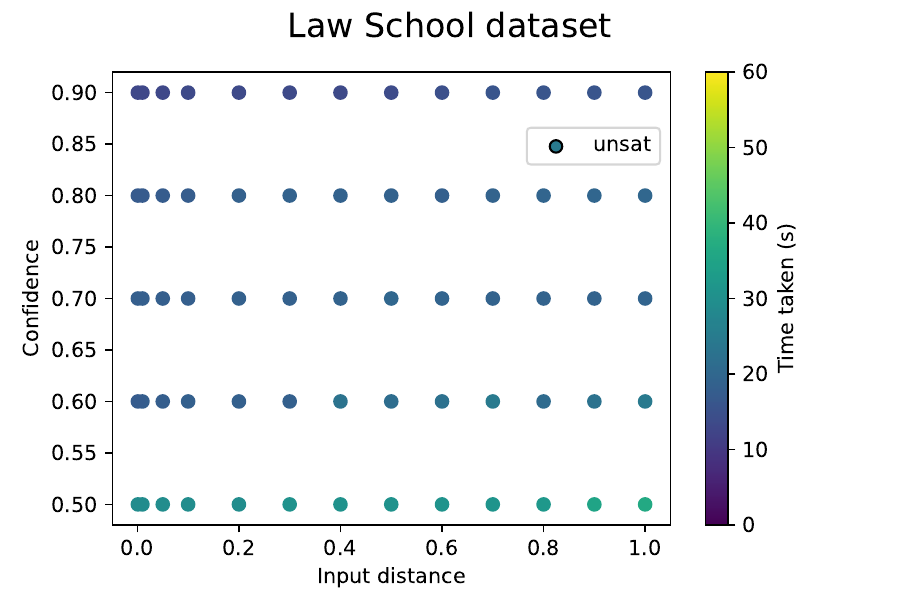} 
		\caption{Input distance vs. confidence for law school dataset} 
            \label{fig:law}
	\end{minipage} 
	\hspace{1.5cm} 
	\begin{minipage}[t]{4cm} 
		\centering 
		\includegraphics[scale=0.4]{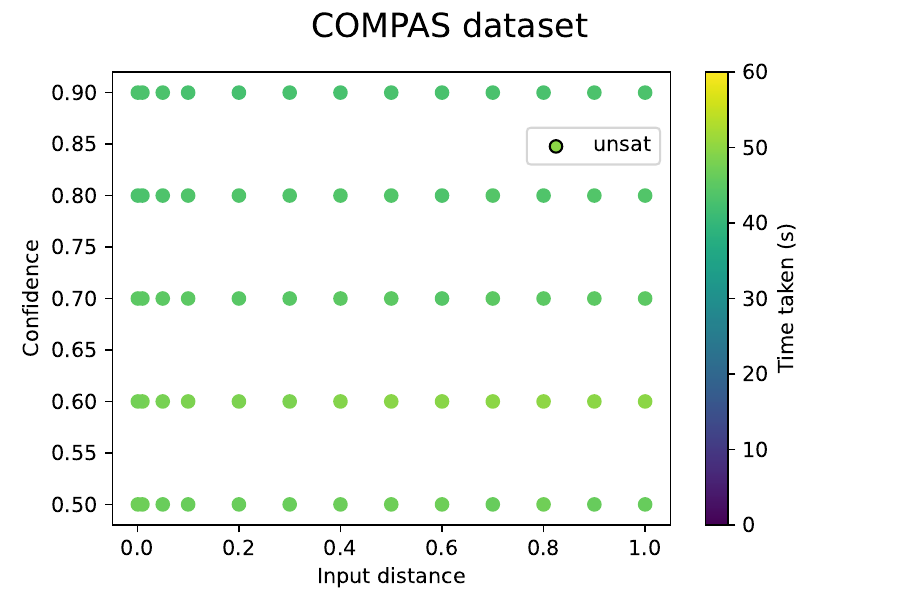} 
		\caption{Input distance vs. confidence for COMPAS dataset} 
            \label{fig:compas}
	\end{minipage} 
        
\end{figure}

First, we present our confidence-based global robustness results. We evaluate our implementation on the neural networks trained with the benchmark datasets for various combinations of input distance and confidence values. We aim to find proofs for globally robust neural networks. The plots in Fig. \ref{fig:german_credit}, \ref{fig:adult}, \ref{fig:law} and \ref{fig:compas} show our experimental results as scatter plots. Markers denoting 'sat' correspond to the query resulting in a counter-example. A counter-example here means that for the input distance and confidence values in that query, the inputs are classified into different output classes. The 'unsat' markers stand for the query being proved (i.e. the model is robust), which means that for the corresponding input distance and confidence threshold, the inputs are classified to the same output class and the model is globally robust. The color bar on the right side denotes the time taken in seconds to run each query; the time scale goes from deep purple to blue, green and yellow as the time taken increases from 0 to 60 sec. 
The plot in \ref{fig:german_credit} depicts the effect of varying input distance and confidence on the German credit benchmark. We ran our query with the confidence-based global robustness property for input distance, $\epsilon$ and confidence $\kappa$ values ranging from 0.001 to 1.0 and 0.5 to 0.9, respectively. Observe that for $\kappa$ values below 0.7, the model is sat i.e. we find counter-examples. However, for confidence values above 0.75, even for larger input distances, the queries result in unsat and  a proof that the model is robust above a confidence threshold of 0.75.

The plots in Figures \ref{fig:adult}, \ref{fig:law} and \ref{fig:compas} show the results for neural networks trained with Adult, Law School, and COMPAS datasets. As can be observed from the scatter plots, these models are robust. For confidence values above 0.5, they are 2-safe and we are successfully able to prove this rather fast in 50 seconds or less.
\vspace{-4ex}
\setlength{\tabcolsep}{2pt}
\begin{table}
    \centering
    \begin{tabular}{ccccc}
        Dataset & Sensitive attribute & Confidence threshold& Result & Time taken  \\
        \hline
        German credit & Gender & 0.5 & unsat & 10.232 sec\\
        German credit & Age & 0.5 & unsat & 11.478 sec\\
        COMPAS & Gender & 0.5 & sat & 7.423 sec\\
        COMPAS & Ethnicity & 0.5 & sat & 18.293 sec\\
        COMPAS & Ethnicity & 0.99 & sat & 25.846 sec\\
        COMPAS & Ethnicity & 0.999 & unsat & 171 min 15 sec\\
         &  &  &  & \\
         &  &  &  & \\
    \end{tabular}
    \vspace{-4ex}
    \caption{Global fairness on German credit/COMPAS datasets for various  criteria}
    \vspace{-5pt}
\label{tab:fairness_results}
\end{table}

Next, we present the results for confidence-based global fairness verification, which are shown in Table \ref{tab:fairness_results}. Each row in the table depicts the verification result for a NN along with the sensitive attribute and confidence threshold considered. If the result is 'unsat', it means that the query is proved (i.e. the model is fair). In other words, for the corresponding sensitive attribute and confidence value constraints, the inputs are classified to the same output class and the model is globally fair. On the other hand, 'sat' corresponds to the query resulting in a counter-example. A counter-example here means that for the corresponding sensitive attribute and confidence threshold in the query, the inputs are classified into different output classes.

The German credit model is proved to be globally fair for confidence values above 0.5 for sensitive attributes Gender and Age. Running our query with the confidence-based global fairness property for the COMPAS model, with Gender as the sensitive attribute gives counter-examples for all confidence values. Additionally, when Ethnicity is considered as the sensitive attribute while verifying the COMPAS model, we find counter-examples for lower confidence values. However, the model is proved to be globally fair for confidence values above 0.999.

We combined our method with binary search, to synthesize the minimum confidence for which the DNN is globally robust or fair. We perform the binary search, starting with confidence 0.5. If the model is unsat, we are done. Else, we check for confidence mid = (0.5 + 1)/2, and continue in this way till we find the minimum confidence accurate to the nearest 0.05. For instance, binary search combined with our method, on German credit gave us 0.75 (in 45 seconds) to be the minimum confidence for which the DNN is globally robust.

Our experimental results on 2-safety properties (with regard to both, global robustness and global fairness), clearly point out that taking confidence along with input distance into account is crucial when verifying neural networks.

\subsection{Discussion}
\subsubsection{Soundness} Our proof of soundness guarantees that if our approach yields that a model is robust or fair for a given confidence and input distance, the model is indeed safe. In case of the German Credit model, for instance, the model is indeed globally robust for all input distances when the confidence is at least 0.75. Moreover, we can use binary search to find the minimal confidence value above which a model is robust.
 Our approach guarantees soundness and when
 a model is found to be safe, it is indeed safe. However, if a counterexample is found, it may be a false positive. 
 False positives (or spurious counterexamples) may in general stem either from overapproximations of the underlying reachability analysis tool or from our own softmax approximation. In our implementation, the former are not present since Marabou is complete (i.e., it does not have false positives), whereas our softmax approximation yields a confidence error that depends on the number of DNN-outputs, as formalized in \autoref{tm} and quantified in its proof. For DNNs with two outputs, such as German Credit, Adult, and Law School, there is no error, whereas for three outputs (COMPAS) the error is $\sim$ 0.171.  Hence, if we want to certify a three-output DNN for confidence $x$, we run our analysis for confidence $x-0.171$: if no attack is found, we can certify the network for confidence $x$ (soundness), otherwise we know the counterexample violates the 2-safety property (completeness of Marabou) for confidence in between $x$ and $x-0.171$ (the possible imprecision is due to our softmax overapproximation). We report counterexamples in this paper on German Credit and on COMPAS: the former are true positives (2-output DNN), whereas the latter are counterexamples for confidence in between $0.999$ (the confidence we can certify the network for) and $0.828$. We ran the network on the counterexample reported by Marabou and found the real confidence to be $0.969$. Hence, the network is for sure fair for a confidence higher than $0.999$, unfair for confidence levels lower than $0.969$, while we cannot decide it for the confidence levels in the interval between $0.969$ and $0.999$. This means that on our datasets, our analysis is very accurate.\subsubsection{Threats to Validity} Presuming a high level of confidence as a precondition can make low-confidence networks vacuously safe. However, an accurate but low-confidence network is not desirable in the first place. This concept is known in the literature as miscalibration. The Expected Calibration Error is defined as weighted average over the absolute difference between confidence and accuracy. In scenarios where accurate confidence measures are crucial, the goal is to reduce the Maximum Calibration Error \cite{naeini2015obtaining} that is the maximum discrepancy between confidence and accuracy.

This is orthogonal to our work and there is an entire field of research \cite{guo2017calibration,ao2023two} aiming at minimizing such calibration errors.

\section{Conclusion}
We introduce the first automated method to verify 2-safety properties such as global robustness and global fairness in DNNs based on the confidence level.  To handle the nonlinear \softmax function computing the confidence, we approximate it with a piece-wise linear function for which we can bound the approximation error. We then compute the self-composition of the DNN with the approximated \softmax and we show how to leverage existing tools such as Marabou to verify 2-safety properties. We prove that our analysis on the approximated network is sound with respect to the original one when the value of confidence is greater than $0.5$ in the approximated one. 
We successfully evaluate our approach on four different DNNs, proving global robustness and global fairness in some cases while finding counterexamples in others.

While we improve over recent verifiers for global properties that are limited to binary classifiers~\cite{BiswasR23}, a limitation of our current approach is that we can only handle DNNs with few (two to five) outputs, since the approximation error increases with the number of outputs. We plan to overcome this limitation in future work by devising more accurate abstractions of \softmax. 

To improve scalability, we will investigate how to refine our approach by integrating  pruning strategies, such as those developed in \cite{BiswasR23},  which we intend to refine to fit our static analysis framework.  

We also plan to explore more sophisticated and effective verification techniques for 2-safety properties, possibly tailored to specific DNN structures. 

Finally, we plan to complement our verification approach with testing techniques to further explore the generated counterexamples.
\section{Acknowledgements}
The work published in this paper is a part of the AI4CSM project that has received funding within the ECSEL JU in collaboration with the European Union’s H2020 Framework Programme (H2020/2014-2020) and National Authorities, under grant agreement No.101007326. This work was also partially supported by the WWTF project ICT22-023, by the WWTF project 10.47379/ICT19018, by the European Research Council
(ERC) under the European Union’s Horizon 2020 research (grant
agreement 771527-BROWSEC), by the Austrian Science Fund (FWF) 10.55776/F85 (project F8510-N); the Vienna Science and
Technology Fund (WWTF) through [ForSmart Grant ID:
10.47379/ICT22007]; the Austrian Research Promotion
Agency (FFG) through the COMET K1 SBA.

%
%
%
%



\bibliographystyle{ieeetr}
\bibliography{refs}

\begin{thebibliography}{10}

\bibitem{gurney1997introduction}
K.~Gurney, {\em An introduction to neural networks}.
\newblock CRC press, 1997.

\bibitem{goodfellow2016deep}
I.~J. Goodfellow, Y.~Bengio, and A.~C. Courville, {\em Deep Learning}.
\newblock Adaptive computation and machine learning, {MIT} Press, 2016.

\bibitem{amato2013artificial}
F.~Amato, A.~L{\'o}pez, E.~M. Pe{\~n}a-M{\'e}ndez, P.~Va{\v{n}}hara, A.~Hampl,
  and J.~Havel, ``Artificial neural networks in medical diagnosis,'' 2013.

\bibitem{rao2018deep}
Q.~Rao and J.~Frtunikj, ``Deep learning for self-driving cars: Chances and
  challenges,'' in {\em Proceedings of the 1st international workshop on
  software engineering for AI in autonomous systems}, pp.~35--38, 2018.

\bibitem{duan2019financial}
J.~Duan, ``Financial system modeling using deep neural networks (dnns) for
  effective risk assessment and prediction,'' {\em Journal of the Franklin
  Institute}, vol.~356, no.~8, pp.~4716--4731, 2019.

\bibitem{TranXJ22}
H.~Tran, W.~Xiang, and T.~T. Johnson, ``Verification approaches for
  learning-enabled autonomous cyber-physical systems,'' {\em {IEEE} Des. Test},
  vol.~39, no.~1, pp.~24--34, 2022.

\bibitem{bjesse2005formal}
P.~Bjesse, ``What is formal verification?,'' {\em ACM SIGDA Newsletter},
  vol.~35, no.~24, pp.~1--es, 2005.

\bibitem{seshia2018formal}
S.~A. Seshia, A.~Desai, T.~Dreossi, D.~J. Fremont, S.~Ghosh, E.~Kim,
  S.~Shivakumar, M.~Vazquez-Chanlatte, and X.~Yue, ``Formal specification for
  deep neural networks,'' in {\em Automated Technology for Verification and
  Analysis: 16th International Symposium, ATVA 2018, Los Angeles, CA, USA,
  October 7-10, 2018, Proceedings 16}, pp.~20--34, Springer, 2018.

\bibitem{katz2017reluplex}
G.~Katz, C.~Barrett, D.~L. Dill, K.~Julian, and M.~J. Kochenderfer, ``Reluplex:
  An efficient smt solver for verifying deep neural networks,'' in {\em
  Computer Aided Verification: 29th International Conference, CAV 2017,
  Heidelberg, Germany, July 24-28, 2017, Proceedings, Part I 30}, pp.~97--117,
  Springer, 2017.

\bibitem{huang2017safety}
X.~Huang, M.~Kwiatkowska, S.~Wang, and M.~Wu, ``Safety verification of deep
  neural networks,'' in {\em Computer Aided Verification: 29th International
  Conference, CAV 2017, Heidelberg, Germany, July 24-28, 2017, Proceedings,
  Part I 30}, pp.~3--29, Springer, 2017.

\bibitem{gopinath2018deepsafe}
D.~Gopinath, G.~Katz, C.~S. P{\u{a}}s{\u{a}}reanu, and C.~Barrett, ``Deepsafe:
  A data-driven approach for assessing robustness of neural networks,'' in {\em
  Automated Technology for Verification and Analysis: 16th International
  Symposium, ATVA 2018, Los Angeles, CA, USA, October 7-10, 2018, Proceedings
  16}, pp.~3--19, Springer, 2018.

\bibitem{urban2020perfectly}
C.~Urban, M.~Christakis, V.~W{\"u}stholz, and F.~Zhang, ``Perfectly parallel
  fairness certification of neural networks,'' {\em Proceedings of the ACM on
  Programming Languages}, vol.~4, no.~OOPSLA, pp.~1--30, 2020.

\bibitem{xie2023deepgemini}
X.~Xie, F.~Zhang, X.~Hu, and L.~Ma, ``Deepgemini: Verifying dependency fairness
  for deep neural network,'' in {\em Proceedings of the AAAI Conference on
  Artificial Intelligence}, vol.~37, pp.~15251--15259, 2023.

\bibitem{pulina2012challenging}
L.~Pulina and A.~Tacchella, ``Challenging smt solvers to verify neural
  networks,'' {\em Ai Communications}, vol.~25, no.~2, pp.~117--135, 2012.

\bibitem{li2019analyzing}
J.~Li, J.~Liu, P.~Yang, L.~Chen, X.~Huang, and L.~Zhang, ``Analyzing deep
  neural networks with symbolic propagation: Towards higher precision and
  faster verification,'' in {\em Static Analysis: 26th International Symposium,
  SAS 2019, Porto, Portugal, October 8--11, 2019, Proceedings 26},
  pp.~296--319, Springer, 2019.

\bibitem{pulina2010abstraction}
L.~Pulina and A.~Tacchella, ``An abstraction-refinement approach to
  verification of artificial neural networks,'' in {\em Computer Aided
  Verification: 22nd International Conference, CAV 2010, Edinburgh, UK, July
  15-19, 2010. Proceedings 22}, pp.~243--257, Springer, 2010.

\bibitem{singh2019abstract}
G.~Singh, T.~Gehr, M.~P{\"u}schel, and M.~Vechev, ``An abstract domain for
  certifying neural networks,'' {\em Proceedings of the ACM on Programming
  Languages}, vol.~3, no.~POPL, pp.~1--30, 2019.

\bibitem{gehr2018ai2}
T.~Gehr, M.~Mirman, D.~Drachsler-Cohen, P.~Tsankov, S.~Chaudhuri, and
  M.~Vechev, ``Ai2: Safety and robustness certification of neural networks with
  abstract interpretation,'' in {\em 2018 IEEE symposium on security and
  privacy (SP)}, pp.~3--18, IEEE, 2018.

\bibitem{10097028}
A.~Baninajjar, K.~Hosseini, A.~Rezine, and A.~Aminifar, ``Safedeep: A scalable
  robustness verification framework for deep neural networks,'' in {\em ICASSP
  2023 - 2023 IEEE International Conference on Acoustics, Speech and Signal
  Processing (ICASSP)}, pp.~1--5, 2023.

\bibitem{cheng2017maximum}
C.-H. Cheng, G.~N{\"u}hrenberg, and H.~Ruess, ``Maximum resilience of
  artificial neural networks,'' in {\em Automated Technology for Verification
  and Analysis: 15th International Symposium, ATVA 2017, Pune, India, October
  3--6, 2017, Proceedings 15}, pp.~251--268, Springer, 2017.

\bibitem{tjeng2017evaluating}
V.~Tjeng, K.~Xiao, and R.~Tedrake, ``Evaluating robustness of neural networks
  with mixed integer programming,'' {\em arXiv preprint arXiv:1711.07356},
  2017.

\bibitem{dutta2018output}
S.~Dutta, S.~Jha, S.~Sankaranarayanan, and A.~Tiwari, ``Output range analysis
  for deep feedforward neural networks,'' in {\em NASA Formal Methods: 10th
  International Symposium, NFM 2018, Newport News, VA, USA, April 17-19, 2018,
  Proceedings 10}, pp.~121--138, Springer, 2018.

\bibitem{katz2019marabou}
G.~Katz, D.~A. Huang, D.~Ibeling, K.~Julian, C.~Lazarus, R.~Lim, P.~Shah,
  S.~Thakoor, H.~Wu, A.~Zelji{\'c}, {\em et~al.}, ``The marabou framework for
  verification and analysis of deep neural networks,'' in {\em Computer Aided
  Verification: 31st International Conference, CAV 2019, New York City, NY,
  USA, July 15-18, 2019, Proceedings, Part I 31}, pp.~443--452, Springer, 2019.

\bibitem{khedr2022certifair}
H.~Khedr and Y.~Shoukry, ``Certifair: A framework for certified global fairness
  of neural networks,'' {\em arXiv preprint arXiv:2205.09927}, 2022.

\bibitem{BiswasR23}
S.~Biswas and H.~Rajan, ``Fairify: Fairness verification of neural networks,''
  in {\em 45th {IEEE/ACM} International Conference on Software Engineering,
  {ICSE} 2023, Melbourne, Australia, May 14-20, 2023}, pp.~1546--1558, {IEEE},
  2023.

\bibitem{chen2021learning}
Y.~Chen, S.~Wang, Y.~Qin, X.~Liao, S.~Jana, and D.~Wagner, ``Learning security
  classifiers with verified global robustness properties,'' in {\em Proceedings
  of the 2021 ACM SIGSAC Conference on Computer and Communications Security},
  pp.~477--494, 2021.

\bibitem{wang2021beta}
S.~Wang, H.~Zhang, K.~Xu, X.~Lin, S.~Jana, C.-J. Hsieh, and J.~Z. Kolter,
  ``Beta-crown: Efficient bound propagation with per-neuron split constraints
  for neural network robustness verification,'' {\em Advances in Neural
  Information Processing Systems}, vol.~34, pp.~29909--29921, 2021.

\bibitem{xu2020fast}
K.~Xu, H.~Zhang, S.~Wang, Y.~Wang, S.~Jana, X.~Lin, and C.-J. Hsieh, ``Fast and
  complete: Enabling complete neural network verification with rapid and
  massively parallel incomplete verifiers,'' {\em arXiv preprint
  arXiv:2011.13824}, 2020.

\bibitem{zhang2018efficient}
H.~Zhang, T.-W. Weng, P.-Y. Chen, C.-J. Hsieh, and L.~Daniel, ``Efficient
  neural network robustness certification with general activation functions,''
  {\em Advances in neural information processing systems}, vol.~31, 2018.

\bibitem{ferrari2022complete}
C.~Ferrari, M.~N. Muller, N.~Jovanovic, and M.~Vechev, ``Complete verification
  via multi-neuron relaxation guided branch-and-bound,'' {\em arXiv preprint
  arXiv:2205.00263}, 2022.

\bibitem{bak2020improved}
S.~Bak, H.-D. Tran, K.~Hobbs, and T.~T. Johnson, ``Improved geometric path
  enumeration for verifying relu neural networks,'' in {\em Computer Aided
  Verification: 32nd International Conference, CAV 2020, Los Angeles, CA, USA,
  July 21--24, 2020, Proceedings, Part I 32}, pp.~66--96, Springer, 2020.

\bibitem{BartheDR11}
G.~Barthe, P.~R. D'Argenio, and T.~Rezk, ``Secure information flow by
  self-composition,'' {\em Math. Struct. Comput. Sci.}, vol.~21, no.~6,
  pp.~1207--1252, 2011.

\bibitem{huang2003learning}
G.-B. Huang, ``Learning capability and storage capacity of two-hidden-layer
  feedforward networks,'' {\em IEEE transactions on neural networks}, vol.~14,
  no.~2, pp.~274--281, 2003.

\bibitem{sharma2017activation}
S.~Sharma, S.~Sharma, and A.~Athaiya, ``Activation functions in neural
  networks,'' {\em Towards Data Sci}, vol.~6, no.~12, pp.~310--316, 2017.

\bibitem{albarghouthi2021introduction}
A.~Albarghouthi {\em et~al.}, ``Introduction to neural network verification,''
  {\em Foundations and Trends{\textregistered} in Programming Languages},
  vol.~7, no.~1--2, pp.~1--157, 2021.

\bibitem{clarkson2010hyperproperties}
M.~R. Clarkson and F.~B. Schneider, ``Hyperproperties,'' {\em Journal of
  Computer Security}, vol.~18, no.~6, pp.~1157--1210, 2010.

\bibitem{barthe2011relational}
G.~Barthe, J.~M. Crespo, and C.~Kunz, ``Relational verification using product
  programs,'' in {\em FM 2011: Formal Methods: 17th International Symposium on
  Formal Methods, Limerick, Ireland, June 20-24, 2011. Proceedings 17},
  pp.~200--214, Springer, 2011.

\bibitem{bastani2016measuring}
O.~Bastani, Y.~Ioannou, L.~Lampropoulos, D.~Vytiniotis, A.~Nori, and
  A.~Criminisi, ``Measuring neural net robustness with constraints,'' {\em
  Advances in neural information processing systems}, vol.~29, 2016.

\bibitem{mangal2019robustness}
R.~Mangal, A.~V. Nori, and A.~Orso, ``Robustness of neural networks: A
  probabilistic and practical approach,'' in {\em 2019 IEEE/ACM 41st
  International Conference on Software Engineering: New Ideas and Emerging
  Results (ICSE-NIER)}, pp.~93--96, IEEE, 2019.

\bibitem{leino2021globally}
K.~Leino, Z.~Wang, and M.~Fredrikson, ``Globally-robust neural networks,'' in
  {\em International Conference on Machine Learning}, pp.~6212--6222, PMLR,
  2021.

\bibitem{binns2018fairness}
R.~Binns, ``Fairness in machine learning: Lessons from political philosophy,''
  in {\em Conference on fairness, accountability and transparency},
  pp.~149--159, PMLR, 2018.

\bibitem{pant2017smooth}
Y.~V. Pant, H.~Abbas, and R.~Mangharam, ``Smooth operator: Control using the
  smooth robustness of temporal logic,'' in {\em 2017 IEEE Conference on
  Control Technology and Applications (CCTA)}, pp.~1235--1240, IEEE, 2017.

\bibitem{remez}
E.~Y. Remez, ``Sur la détermination des polynômes d'approximation de degré
  donnée,'' 1934.

\bibitem{hofmann1994german}
H.~Hofmann, ``German credit dataset,'' {\em UCI Machine 2023 Repository
  [http://archive. ics. uci. edu/ml]. University of California, School of
  Information and Computer Science, Irvine}, 1994.

\bibitem{Dua:2019}
D.~Dua and C.~Graff, ``{UCI} machine 2023 repository,'' 2017.

\bibitem{compas}
J.~Larson. {https://github.com/propublica/compas-analysis}, 2017.

\bibitem{wightman1998lsac}
L.~F. Wightman, ``Lsac national longitudinal bar passage study. lsac research
  report series.,'' 1998.

\bibitem{naeini2015obtaining}
M.~P. Naeini, G.~Cooper, and M.~Hauskrecht, ``Obtaining well calibrated
  probabilities using bayesian binning,'' in {\em Proceedings of the AAAI
  conference on artificial intelligence}, vol.~29, 2015.

\bibitem{guo2017calibration}
C.~Guo, G.~Pleiss, Y.~Sun, and K.~Q. Weinberger, ``On calibration of modern
  neural networks,'' in {\em International conference on machine learning},
  pp.~1321--1330, PMLR, 2017.

\bibitem{ao2023two}
S.~Ao, S.~Rueger, and A.~Siddharthan, ``Two sides of miscalibration:
  identifying over and under-confidence prediction for network calibration,''
  in {\em Uncertainty in Artificial Intelligence}, pp.~77--87, PMLR, 2023.

\end{thebibliography}

\clearpage
\appendix
\section{Appendix}
\label{sec:appendix}
\begin{manualtheorem}{1}
Let $\softmax$ and $\widehat{\softmax}$ compute the real and linearly approximated softmax (with precision $\delta$), respectively for the last layer of $n \geq 2$ neurons $\vec{z}$ of a neural network and $z_i = max(z_1, \cdots, z_n)$. Then, we have the following result: 
   \begin{equation*}
   \label{eq:conf}
   \forall \vec{z} . \ \softmax(\vec{z})\textsubscript i - \widehat{\softmax}(\vec{z}) \textsubscript i \leq \frac{n-2}{(\sqrt{n-1} + 1)^2} + 2 \delta 
   \end{equation*}
\end{manualtheorem}

\begin{proof} 
The error bound is the difference between the upper and lower bounds of the $\softmax$ and it is a function of the number of outputs $n$ of the neural network and the error $\delta$ for the approximation of the sigmoid.

\begin{equation*}
\begin{split}
     \gamma(n,\delta,\vec{z})\textsubscript i =
     \Sigmoidapp(z{\textsubscript{\tiny i}} -  \mathop{max_{1}^{n}}_{j\neq i}(z_j)) + \delta -  \widehat{\softmax}  (\vec{z})\textsubscript i 
\end{split}
\end{equation*}
\begin{equation*}
\begin{split}
     \gamma(n,\delta,\vec{z})\textsubscript i = \frac{\alpha(\vec{z}) \textsubscript i  (n -2)}{(\alpha (\vec{z})\textsubscript i +1) ((n-1)\alpha(\vec{z})\textsubscript i + 1)} + 2\delta 
\end{split}
\end{equation*}
where: 
\begin{equation*}
\begin{split}
     \alpha(\vec{z}) \textsubscript i  = e^{-z_i + \mathop{max_{1}^{n}}_{j\neq i}(z_j)}
\end{split}
\end{equation*}

Since $z_i = max(z_1, \cdots, z_n)$ then is always $0 < \alpha(\vec{z}) \textsubscript i  \leq 1 $. 
Let us now replace $\alpha(\vec{z}) \textsubscript i $ with the variable $m$ and study the function in the domain $m \in ]0,1]$ with $n \in \mathbb{N}, n \geq 2$:
\begin{equation*}
\begin{split}
\zeta_n(m) = \frac{m  (n -2)}{((m +1) ((n-1)m + 1)}
\end{split}
\end{equation*}
To compute the maximum in the domain $m \in ]0,1]$ we first compute the derivative using the quotient rule:
\begin{equation*}
\begin{split}
\zeta'_n(m) = \frac{(n -2) ((1-n)m^2 + 1)}{((m +1)^2 ((n-1)m + 1)^2}
\end{split}
\end{equation*}
To find the point $m_{max}$ for which we have the maximum $\zeta_n(m_{max})$, we need to find the positive root for the following equation:
\begin{equation*}
\begin{split}
(n -2) ((1-n)m^2 + 1) = 0
\end{split}
\end{equation*}
The positive root is: 
\begin{equation*}
\begin{split}
m_{max} = \frac{1}{\sqrt{n-1}}
\end{split}
\end{equation*}
As a consequence, the maximum is:
\begin{equation*}
\begin{split}
\zeta_n(m_{max}) = \frac{n-2}{(\sqrt{n-1} + 1)^2} 
\end{split}
\end{equation*}
If the number of outputs is two, the softmax behaves as a sigmoid. Thus, the only error is the one due to the piece-wise approximation ($2 \delta$) of the sigmoid. Let us assume to have three outputs and $\delta=0.0001$ then the maximum error between the real softmax and the approximated one is $\frac{1}{(\sqrt{2} + 1)^2} + 0.0002 \sim 0.17177$.

\end{proof}

\begin{manualtheorem}{2} (Class consistency)
 Let $f$ and $\hat{f}$ denote the real and the approximated (with precision $\delta$) neural networks with $n\geq 2$ outputs, respectively. Then:
 $$\conf(\hat{f}(\vec{x})) > \frac{1}{2} \implies \class(\hat{f}(\vec{x})) = \class(f(\vec{x})) $$
 \end{manualtheorem}

\begin{proof} 
    $\conf(\hat{f}(\vec{x}))=\max(\hat{\out}(v_{l,1}),\ldots,\hat{\out}(v_{l,n})) > \frac{1}{2}$. 
    Let us assume that the component $\hat{\out}(v_{l,i}) > \frac{1}{2}$ is the maximum value resulting from applying the approximated softmax. From Theorem 1 we have that $\out(v_{l,i}) \geq \hat{\out}(v_{l,i}) > \frac{1}{2}$ and consequently $i$ is the class for both the real and the approximated neural network $\class(f(\vec{x})) = \class(\hat{f}(\vec{x}))$.
\end{proof}

\begin{manualtheorem} {3} (Soundness) Let $f$ and $\hat{f}$ be the original neural network and over-approximated neural network, respectively. Let $b_{n,\delta}$ be 
the error bound of the approximated softmax ($b_{n,\delta} =\frac{n-2}{(\sqrt{n-1} + 1)^2} + 2 \delta$ (see Theorem \ref{tm})). Then we have the following soundness guarantee: Whenever the approximated neural network is 2-safe for $\conf(\hat{f}(\vec{x})) > (\kappa - b_{n,\delta})$, the real neural network is 2-safe for $\conf(f(\vec{x})) > \kappa$, given $\conf(\hat{f}(\vec{x})) > \frac{1}{2}$. Formally:
\begin{multline*} 
 \left(\begin{split}
 \forall \vec{x}, \vec{x'}. \cond(\vec{x}, \vec{x'}, \vec{\epsilon}) \land {\conf}(\hat{f}(\vec{x})) > (\kappa - b_{n,\delta}) \\ \implies {\class}(\hat{f}(\vec{x})) = {\class}(\hat{f}(\vec{x'})) 
\end{split}\right) \implies \\
\left(\begin{split}\forall \vec{x}, \vec{x'}. \ \cond(\vec{x}, \vec{x'}, \vec{\epsilon}) \land \conf(\vec{f(x)}) > \kappa  \\ \implies \class(f(\vec{x})) = \class(f(\vec{x'}))\end{split}\right),
\mbox{ with } \: \conf(\hat{f}(\vec{x})) > \frac{1}{2}
\end{multline*} 
\end{manualtheorem}

\begin{proof}
    
Let us consider all possible cases around the implications one by one and see if they hold. The theorem to prove is of the form $(\hat{A} \implies \hat{B}) \implies (A \implies B)$ where:
\begin{align*}
&\hat{A} \defn  \left(\forall \vec{x}, \vec{x'}.\cond(\vec{x}, \vec{x'}, \vec{\epsilon}) \land {\conf}(\hat{f}(\vec{x})) > (\kappa - b_{n,\delta})\right), \\
 &\hat{B} \defn \left({\class}(\hat{f}(\vec{x})) = {\class}(\hat{f}(\vec{x'}))\right) ,\\
&A \defn  \left(\forall \vec{x}, \vec{x'}. \cond(\vec{x}, \vec{x'}, \vec{\epsilon}) \land \conf(\vec{f(x)}) > \kappa\right), \\
&B \defn  \left(\class(f(\vec{x})) = \class(f(\vec{x'}))\right) 
\end{align*}

Case 1: If ${\conf}(\hat{f}(\vec{x})) \leq \kappa - b_{n,\delta}$ then $\conf(f(\vec{x})) \leq \kappa$ follows from Theorem 1 and the assumption that $\conf(\hat{f}(\vec{x})) > \frac{1}{2}$. $\hat{A}$ will be false and $\neg \hat{A} \implies \neg A$ so also $A$ is false and 
 $(\hat{A} \implies \hat{B}) \implies (A \implies B)$ is true. \\

Case 2: Assume ${\conf}(\hat{f}(\vec{x})) > \kappa - b_{n,\delta}$. From Theorem 2 we have that $\class(\hat{f}(\vec{x})) = \class(f(\vec{x}))$ and $\class(\hat{f}(\vec{x'})) = \class(f(\vec{x'}))$ so $\hat{B} \implies B$ and then independently of $A$ we have that $(\hat{A} \implies \hat{B}) \implies (A \implies B)$ is true.



The soundness theorem, thereby, states that if the confidence-based 2-safety property holds on the over-approximated neural network, it also holds on the original network. 
\end{proof}


\end{document}